# Influence of grain size and grain boundary misorientation on the fatigue crack initiation mechanisms of textured AZ31 Mg alloy

Abbas Jamali[a,b], Anxin Ma[a], Javier LLorca[a,b],*

[a] *IMDEA Materials Institute, Getafe, Madrid 28906, Spain*
[b] *Department of Materials Science, Polytechnic University of Madrid/Universidad Politécnica de Madrid, 28040 Madrid, Spain*

*Corresponding author: javier.llorca@upm.es, javier.llorca@imdea.org

## Abstract

The deformation and crack initiation mechanisms were analyzed in a textured AZ31B-O Mg alloy subjected to fully-reversed, strain-controlled cyclic deformation along the rolling direction after 50 cycles (approximately 33% of the fatigue life). Distinct deformation bands corresponding to pyramidal slip or tensile twins were found in 538 grains out of 2100 grains. Slip trace analysis showed that 72.3% were pyramidal slip bands and 18.4% were twin boundaries. Both pyramidal slip and twinning was only found in 9.1% of the grains with deformation bands. Cracking was widespread after 50 cycles. Grain boundary cracks were found in ≈15% of the small grains (< 20 μm) and they were mainly associated with high angle grain boundaries (>40º). Cracking was also found to occur by transgranular cracks parallel to the pyramidal slip bands or twin boundaries in large grains (>45 μm). The majority (>60%) of these large grains presented transgranular cracks after 50 cycles.

Mg alloys present large potential for structural applications in transport (aerospace, automotive) because of their low density and good castability together with reasonable stiffness and strength [1,2]. Furthermore, Mg alloys are also appealing -among biodegradable metals- for bone implants because of its biocompatibility and osteopromotive properties, which can stimulate new bone formation [3,4]. Cyclic loads are always present in these applications and, thus, the fatigue behavior of Mg alloys has been studied by different authors [5–19]. Wrought Mg alloys usually had a strong basal texture along the rolling direction which, leads to a very large tension-compression asymmetry during cyclic loading due to the large differences in the critical resolved shear stress (CRSS) between basal and pyramidal slip and the activation of tensile twinning [5,6,12,13]. Compression deformation along the rolling direction (RD) has to be accommodated by tensile twinning because the Schmid factor for basal slip is very low and the CRSS for pyramidal slip is too high. Upon unloading, the reverse strain is initially absorbed by de-twinning but pyramidal slip becomes the dominant deformation mechanism after full de-twinning. Because the CRSS for pyramidal slip is much higher than that for twinning, the peak stress in compression is much lower than that in tension under fully-reversed strain-controlled cyclic deformation. The opposite behavior is found when the textured Mg alloys are deformed along the normal direction (ND), with twining being the dominant deformation mechanism in tension and de-twinning followed by pyramidal slip in compression [14,15]. The tension-compression asymmetry during cyclic deformation increases with the applied cyclic strain amplitude and with the texture because both enhance the activation of tensile twinning. However, it is reduced in the case of weakly textured alloys [16] or if the loading direction is favorable for basal slip because most of the cyclic plastic strain in these cases can be accommodated by this mechanism due to low CRSS to activate basal slip [18,19].

The differences in the cyclic deformation mechanisms lead to important differences in the fatigue life as a function of orientation and of the cyclic strain amplitude [18,19], as opposed to other metallic alloys which show an isotropic mechanical response in fatigue. These differences are associated with changes in the damage accumulation mechanisms in the microstructure which eventually lead to the initiation of fatigue cracks. Different sites for fatigue crack initiation have been reported in Mg alloys, including intermetallic particles [8], grain boundaries [7,14] persistent slip bands [7,8,14] and transgranular cracks associated with twinning/de-twinning bands [8–10,14]. However, statistical studies that indicate which one is dominant are lacking. This information is critical to develop fatigue indicator parameters that can be used -in combination with computational homogenization and crystal plasticity- to predict the fatigue life of Mg alloys [11,17,20] and to design microstructures with improved fatigue resistance.

In this investigation, the fatigue crack initiation mechanisms under fully-reversed cyclic deformation were investigated in a rolled AZ31 magnesium alloy deformed along the rolling direction. Slabs of 80×65×500 mm$^3$ of a AZ31B-O Mg alloy were purchased from Magnesium Elektron. The nominal chemical composition of the alloy contained 2.89 wt.% Al, 1.05 wt.% Zn and 0.42 wt.% Mn. Flat dogbone samples with a gauge length of 12.5 mm and square cross section of 6 × 6 mm$^2$ were fabricated by electrical discharge machining from the slab with the longest dimension of the sample parallel to the rolling direction. The surfaces of the samples were manually ground on consecutive abrasive SiC papers with a grit size of 1200, 2000 and 4000 followed by three polishing steps with 3 μm, 1 μm, and 0.25 μm diamond paste. After polishing, they were cleaned by immersion in an ultrasounds bath of pure ethanol during 3 minutes and etched with a solution containing 0.4 g of picric acid, 2.5 ml of acetic acid, 2 ml of distilled H$_2$O and 4 ml of ethanol to reveal the grain boundaries. Then, they were again etched with Nital 15% (15% nitric acid and 85% ethanol) to get a mirror-like surface. The microstructural features of the alloy were analyzed using a FEI Helios Nanolab 600i scanning electron microscope (SEM) equipped with a Nordlys electron backscattered diffraction (EBSD) detector from Oxford Instruments. Statistically representative maps containing about 2000 grains were collected at an acceleration voltage of 20 kV and acceleration current of 2.7 nA. The step size for the EBSD measurement was one-twelfth of the average grain size to guarantee an acceptable mapping resolution. The grain size distribution and crystal orientation were extracted from the raw data using the MTEX software [21].

Fatigue tests were carried out under strain control under fully reversed deformation at a cyclic strain semi-amplitude of $\Delta\varepsilon/2 = 2.0\%$ in a servo-hydraulic Instron 8802 mechanical testing machine. This strain was selected because it led to large number of microcracks on the specimen surface after 50 cycles, which allowed to relate the crack initiation sites with the microstructural features. Strain was controlled with an extensometer attached to the sample, the test frequency was 0.5 Hz and load and deformation were monitored during the test through a computer-controlled data acquisition system. Specimens were initially deformed in compression in all cases. Three tests were carried out until failure while two tests were stopped after 5 and 50 cycles at the minimum (compression) strain. They were removed from the machine and the sample surfaces were observed in the SEM using EBSD and secondary electron detectors to assess the deformation mechanisms and the fatigue crack initiation sites.

The microstructure of the rolled AZ31B-O alloy in the RD-TD plane is shown in the Fig.1a. It shows a very strong basal texture with the basal plane of most grains parallel to the RD-TD plane. The analysis of the grain boundary misorientation distribution showed the presence of very few twinned grains (Fig.1c, d), which showed different orientation and that are marked with black arrows in Fig. 1a. The grain size distribution is plotted in Fig. 1b and the average grain size was 14.4 ± 8.8 µm. Approximately 80% of the grains were < 20 µm but the 20% remaining were much larger, reaching up to 60 µm. The average aspect ratio was 1.7 ± 0.8 and the grains (particularly the large ones) were slightly elongated along RD. The grain boundary misorientation is depicted in Fig. 1c, and the corresponding misorientation angle distribution is plotted in Fig. 1d. Most of the grain boundaries (≈ 75%) have a misorientation angle <40º. About 2% of the grain boundaries have a misorientation angle between 80º and 90º, which corresponds to the twin boundary misorientation in Mg magnesium alloys (86º).

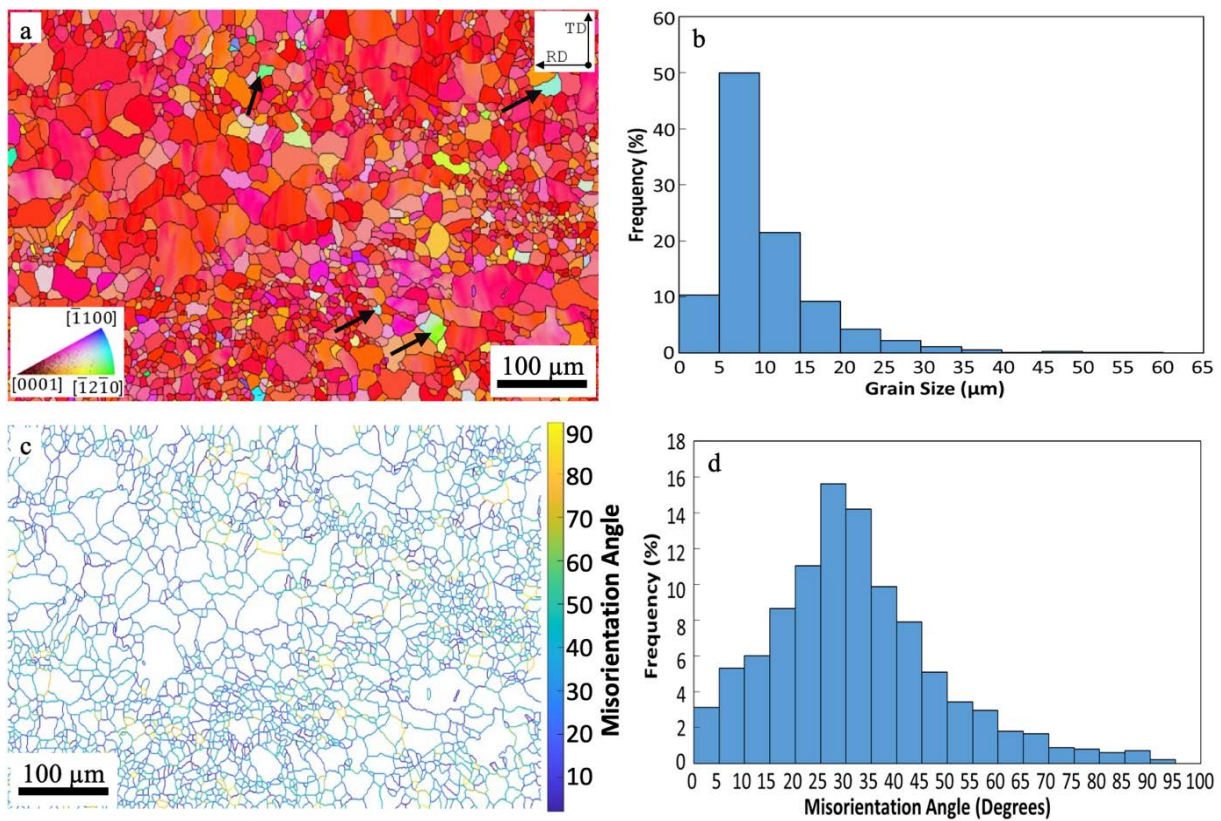

**Fig. 1.** (a) EBSD image showing the orientation of the grains in RD-TD plane. Twinned grains (as indicated by the orientation and the grain boundary misorientation angle) are marked with arrows) (b) Grain size distribution in the RD-TD plane. It should be noted that a few grains in the range 50 to 60 µm were found in the microstructure. (c) Map of the grain boundary misorientation angle. (d) Grain boundary misorientation angle distribution.

The cyclic stress-strain curves of one sample after 1, 2, 10 and 100 fatigue cycles are plotted in Fig. 2a and they are typical of rolled Mg alloys with strong texture. Plastic deformation in the first compression cycle occurs at a constant stress due to twinning and reversed deformation leads to a curve with sigmoidal shape due to the activation de-twinning followed by pyramidal slip. In addition, basal slip is also assumed to develop during the tensile and compressive parts of the fatigue cycle due to the low CRSS necessary to activate this mechanism [5,6,12–15]. The shape of the cyclic stress-strain curves becomes stable after the first cycle, leading to large differences between the maximum stress in tension (controlled by pyramidal slip) and the minimum stress in compression (controlled by twinning). The evolution of the maximum and minimum stresses in the three samples that were deformed until failure is plotted in Fig. 2b. The maximum tensile stress remained constant during the whole fatigue life while continuous hardening was observed in the case of the minimum stress. The three samples failed at approximately 150 cycles. Cyclic hardening in compression was attributed to the presence of residual twins [6,12,19] that are obstacles to dislocation slip while the cyclic hardening in tension due to dislocation accumulation is balanced by the development of damage, as shown below.

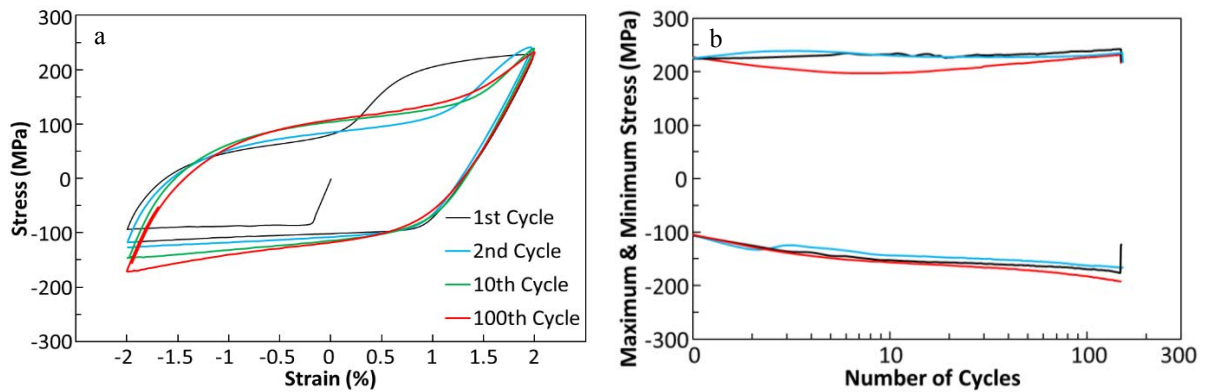

**Fig. 2.** (a) Cyclic stress-strain curves along the RD specimen after 1, 2, 10 and 100 fatigue cycles. The applied cyclic strain semi-amplitude was $\Delta\varepsilon/2 = 2.0\%$. (b) Evolution of the maximum and minimum stress in each cycle as a function of the number of cycles.

The surface of the specimen deformed during 5 cycles was analyzed in the SEM and cracks were not found, although deformation bands were present in many grains. The EBSD map of the sample after 50 cycles is plotted in Fig. 3a. The test was stopped at the minimum strain and the inverse pole figure map shows that the $c$ axis of several grains was reoriented nearly 90° from ND towards RD and the $<11\bar{2}0>$ or $<10\bar{1}0>$ axes were parallel to ND. This texture change is consistent with the activation of tension twins. In addition, although partial de-twinning took place during unloading [15], many elongated twin bands are visible in the microstructure. They were also identified as tension twins because the orientation with respect to the parent grain as well as the grain boundary misorientation. The twin area fraction was 28%, indicating the importance of the twinning/de-twinning mechanism during cyclic deformation.

The deformation mechanisms after 50 fatigue cycles were analyzed from secondary electron images at higher magnification in 2100 grains (Fig. 3b). Distinct deformation features could be identified in 538 grains, which were identified as pyramidal slip bands or tension twins. Most grains contained one set of long and straight slip lines parallel to each other. Slip trace analysis from the grain orientation, following the methodology presented in [22-23], showed that they were parallel to one of the pyramidal slip systems (normally the one with highest Schmid factor). They are indicated by green lines in Fig. 3b. Another set of deformation bands were parallel to one of the tension twin planes - as indicated by the twin trace analysis- and they are identified by the red lines in Fig. 3b. In addition, some small grains appeared brighter in the secondary electron images (marked with orange arrows in Fig. 3b) and they were identified as fully twinned grains by checking with the EBSD map after deformation. The brighter appearance in the secondary electron images results because they were slightly elevated above the sample surface to accommodate the eigenstrain associated with tension twinning. It should be noted that only 11 small grains (with a grain size below 20 μm) out of approximately 2000 grains were fully twinned. Thus, twinning occurred preferentially in large grains and fully-twinned small grains were very rare. Twinning of small grains was very likely triggered by the large local stresses induced by the surrounding grains on these small grains. Finally, basal slip traces were only found in one grain. They were much thinner that the pyramidal slip bands and more difficult to identify. The fraction of each type of features found in the 538 grains is reported in Table 1. It should be noted that only one deformation mechanism (either pyramidal slip or tensile twinning) was found in > 90% of the grains with distinct deformation features and a combination of both of them was reported in 9.1%. Overall, the deformation mechanisms were in agreement with the shape of the stress-strain curves in Fig. 2 and with those reported by other investigations in extruded Mg alloys [5,6,12–17]. The lack of evidence for basal slip in the surfaces does not rule out that this mechanism is also active because the basal slip plane is parallel to the RD-TD plane in most grains and basal slip leaves no traces on this surface.

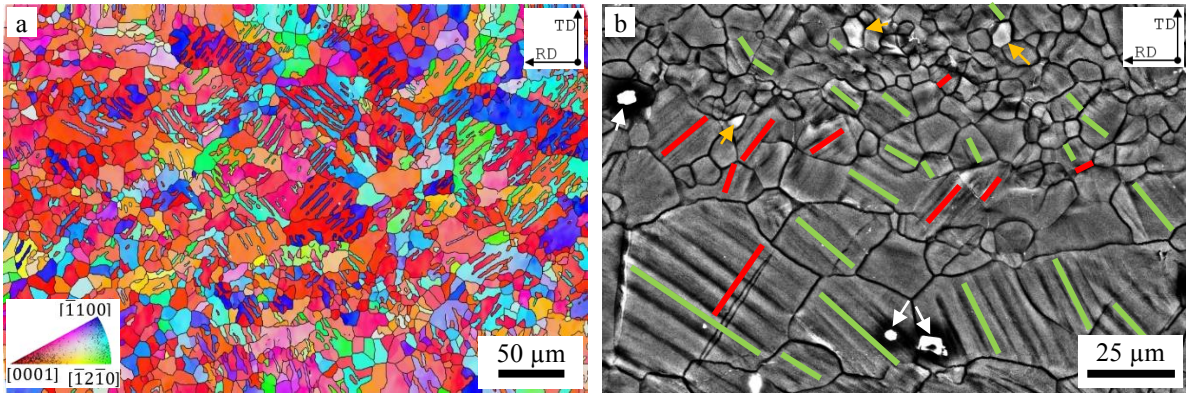

**Fig. 3.** Microstructure of the AZ31B-O sample after 50 fatigue cycles at Δε/2 = 2.0%. (a) EBSD map showing the orientation of the grains and the twins within the grains. (b) Secondary electron image showing deformation bands within the grains. Following slip/twin trace analysis, those parallel to the green lines were identified as pyramidal slip bands and those parallel to the red lines are twins. Intermetallic particles are marked with white arrows and fully twinned small grains with orange arrows.

**Table 1.** Deformation mechanisms observed in 538 grains after 50 fatigue cycles.

| Deformation mechanisms | |
|---|---|
| Pyramidal slip | 72.3% |
| Tensile twinning | 18.4% |
| Pyramidal slip & tensile twinning | 9.1% |
| Basal slip | 0.2% |

Fatigue crack initiation sites were analyzed in the same region in which the deformation mechanisms were ascertained by means of higher magnification secondary electron images. A total of 496 microcracks were found, indicating that microcracking was widespread after 33% of the fatigue life. Transgranular and intergranular cracks were found (Fig. 4a to 4d), in agreement with previous investigations [7–10,14], together with a few cracks associated with intermetallic particles. The fraction of each type of cracks is detailed in Table 2, which shows that most of the cracks were nucleated at grain boundaries or within slip bands corresponding to pyramidal slip. Cracks at tensile twin boundaries were also found but the fraction was much smaller while cracks rarely were nucleated at intermetallic particles. Crack formation along persistent slip bands is a classical fatigue crack initiation mechanisms [24] while cleavage along twin boundaries has been found in Mg alloys subjected to fatigue [8–10,14]. However, it is interesting to notice that > 60% of the grains with an average grain size above 45 µm were cracked after 50 cycles (Fig. 4e) and that fatigue damage initiation in these large grains was always associated to transgranular cracks. Although there were very few grains larger than 45 µm in the material (Fig. 1b), pyramidal slip and twinning occur more easily in these large grains, leading to localization of the deformation and to the nucleation of large fatigue cracks that encompass the whole grain width. As a result, these transgranular cracks nucleated at the largest grains may become the critical ones that control the fatigue life. This result is in agreement with experimental results of the negative effect of abnormal grain growth on the fatigue endurance of Mg alloys [25] and supports previous predictions of the fatigue life of Mg alloys based on fatigue indicator parameters associated with the shear strain accumulated by plastic slip [17] or with twinning/de-twinning [11] in each fatigue cycle.

**Table 2.** Distribution of 496 fatigue cracks after 50 fatigue cycles.

| Crack initiation site | |
|---|---|
| Grain boundary | 57.2% |
| Pyramidal slip band | 33.9% |
| Tensile twin boundary | 8.1% |
| Intermetallic particle | 0.8% |

Intergranular and transgranular cracks were found in approximately 15 % of the grains < 20 µm (Fig. 4e) but grain boundary cracking was dominant for small grains. Two different types of grain boundary cracks were found. The first one was associated with small, fully twinned grains (Fig. 4c) and the grain boundary cracks appeared due to the stress concentration induced by the twinning eigenstrain. However, this mechanism was only observed in 11 grains. Most of the grain boundary cracks were not associated with twinning neither with the interaction of slip bands or twins with the grain boundary but with large values of the grain boundary misorientation angle, as shown in Fig. 4f. In fact, the fraction of grain boundaries that nucleate fatigue cracks rises sharply when the grain boundary misorientations is > 40º and 70% of the grain boundaries with a misorientation angle between 65º and 90º were cracked. On the contrary, only about 3% of the grain boundaries with the misorientation angle of < 40º were cracked. These cracks in grain boundaries with high misorientation are probably created due to the incompatibility of deformation between adjacent grains, which is more likely to occur in small grains because the critical stress to nucleate twins or promote slip increases as the grain size decreases. As a result, higher stresses are attained at the grain boundaries, promoting grain boundary cracking.

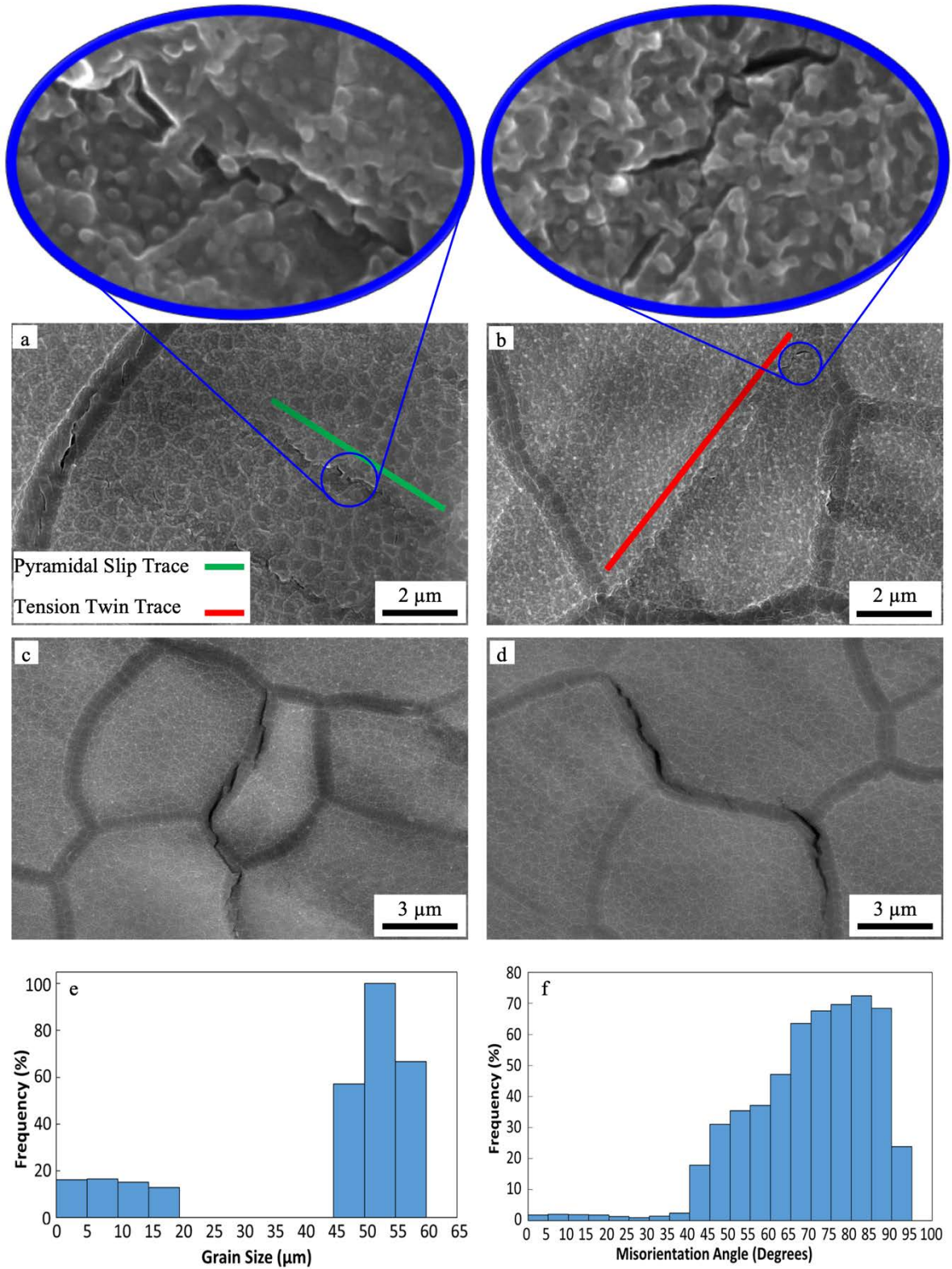

**Fig. 4.** Fatigue crack initiation sites in the extruded AZ31B-O sample after 50 fatigue cycles at Δε/2 = 2.0%. (a) Transgranular cracks parallel to pyramidal slip bands. (b) Transgranular crack parallel to tensile twin. (c) Intergranular crack around a fully twinned grain. (d) Intergranular crack. (e) Fraction of crack initiation sites as a function of the grain size. Fatigue cracks in grains ≥ 50 μm were transgranular, either associated with pyramidal slip

bands or twin boundaries. Fatigue cracks in grains < 25 μm were either intergranular or transgranular. (f) Fraction of cracked grain boundaries as a function of the misorientation angle.

In summary, fatigue deformation and crack initiation mechanisms under fully-reversed, strain-controlled cyclic deformation were studied in a textured AZ31B-O alloy by means of the analysis of the sample surface after 5 and 50 cycles. Plastic deformation during fatigue was accommodated by twinning in the compressive part of the loading cycle, and reversed deformation was accompanied by the activation of de-twinning followed by pyramidal slip in tension once de-twinning was exhausted. These deformation mechanisms were corroborated by the presence of pyramidal slip bands and twin traces in the sample surface, which also showed that only one deformation mechanism (either pyramidal slip or twinning) was dominant in most grains. No cracks were found after 5 fatigue cycles but cracking was widespread after 50 cycles (33% of the fatigue life). Approximately 15% of the small grains (< 20 μm) in the microstructure presented grain boundary cracks. A few of them were associated with the stress concentrations induced by fully twinned grains but the large majority was associated with grain boundaries with large misorientation angle (> 40º). In fact, 70% of the grain boundaries with a misorientation angle between 65º and 90º were cracked because the incompatibility of deformation between neighbor grains could not be accommodated by plastic slip. Cracking was also found to occur by transgranular cracks parallel to the pyramidal slip bands or twin boundaries in large grains (>45 μm). More than 60% of these large grains showed transgranular cracks after 50 cycles. Thus, cracking of grain boundaries with large misorientation angle and transgranular cracking in large grains stand for the dominant deformation mechanisms in fatigue in textured AZ31B-O alloy. This information is important to design new Mg alloys with improved fatigue life and also to develop fatigue life prediction tools based on fatigue indicator parameters.

Declaration of Competing Interest

The authors declare that they have no known competing financial interests or personal relationships that could have appeared to influence the work reported in this paper.

Acknowledgements

This investigation was supported by the European Union Horizon 2020 research and innovation programme (Marie Sklodowska-Curie Individual Fellowships, Grant Agreement 795658) and the Comunidad de Madrid Talento-Mod1 programme (Grant Agreement PR-00096). Additional support by the HexaGB project of the Spanish Ministry of Science (reference RTI2018-098245) is also gratefully acknowledged.

**References**

[1] B.L. Mordike, T. Ebert, Mater. Sci. Eng. A 302 (2001) 37–45.

[2] H. Miura, T. Maruoka, X. Yang, J.J. Jonas, Scr. Mater. 66 (2012) 49–51.

[3] H. Amano, K. Hanada, A. Hinoki, T. Tainaka, C. Shirota, W. Sumida, K. Yokota, N. Murase, K. Oshima, K. Chiba, Y. Tanaka, H. Uchida, Sci. Rep. 9 (2019) 1–8.

[4] L. Mao, L. Shen, J. Chen, X. Zhang, M. Kwak, Y. Wu, R. Fan, L. Zhang, J. Pei, G. Yuan, C. Song, J. Ge, W. Ding, Sci. Rep. 7 (2017) 1–12.

[5] L. Wu, A. Jain, D.W. Brown, G.M. Stoica, S.R. Agnew, B. Clausen, D.E. Fielden, P.K. Liaw, Acta Mater. 56 (2008) 688–695.

[6] S.H. Park, S.G. Hong, W. Bang, C.S. Lee, Mater. Sci. Eng. A 527 (2010) 417–423.

[7] F. Wang, J. Dong, M. Feng, J. Sun, W. Ding, Y. Jiang, Mater. Sci. Eng. A 589 (2014) 209–216.

[8] B. Wen, F. Wang, L. Jin, J. Dong, Mater. Sci. Eng. A 667 (2016) 171–178.

[9] F. Yang, S.M. Yin, S.X. Li, Z.F. Zhang, Mater. Sci. Eng. A 491 (2008) 131–136.

[10] Z. Wang, S. Wu, G. Kang, H. Li, Z. Wu, Y. Fu, P.J. Withers, Acta Mater. 211 (2021) 116881.

[11] F. Briffod, T. Shiraiwa, M. Enoki, Mater. Sci. Eng. A 753 (2019) 79–90.

[12] Y. Xiong, Q. Yu, Y. Jiang, Int. J. Plast. 53 (2014) 107–124.

[13] J.B. Jordon, H.R. Brown, H. El Kadiri, H.M. Kistler, R.L. Lett, J.C. Baird, A.A. Luo, Int. J. Fatigue 51 (2013) 8–14.

[14] Q. Yu, J. Zhang, Y. Jiang, Mater. Sci. Eng. A 528 (2011) 7816–7826.

[15] A.D. Murphy-Leonard, D.C. Pagan, A. Beaudoin, M.P. Miller, J.E. Allison, Int. J. Fatigue 125 (2019) 314–323.

[16] F. Wang, J. Dong, Y. Jiang, W. Ding, Mater. Sci. Eng. A 561 (2013) 403–410.

[17] M. Zhang, H. Zhang, A. Ma, J. Llorca, Int. J. Plast. 139 (2021) 102885.

[18] Y. Xiong, Y. Jiang, Mater. Sci. Eng. A 677 (2016) 58–67.

[19] Y. Wang, D. Culbertson, Y. Jiang, Mater. Des. 186 (2020) 108266.

[20] J. Segurado, R.A. Lebensohn, J. Llorca, Adv. Appl. Mech. 51 (2018) 1-114.

[21] F. Bachmann, R. Hielscher, H. Schaeben, Solid State Phenom. 160 (2010) 63–68.

[22] T.R. Bieler, R. Alizadeh, M. Peña-Ortega, J. LLorca, Int. J. Plast. 118 (2019) 269–290.

[23] R. Alizadeh, M. Peña-Ortega, T. R. Bieler, J. LLorca. Scr. Mater. 78 (2020) 408-412.

[24] S. Suresh, Fatigue of Materials, 2$^{nd}$ edition, Cambridge University Press, 1998.

[25] M.A. Azeem, A. Tewari, U. Ramamurty, Mater. Sci. Engng. A 527 (2010) 898-903.